\documentclass[letterpaper]{article}
\usepackage{STY/aaai}
\usepackage{times}
\usepackage{helvet}
\usepackage{courier}
\usepackage{graphicx}
\usepackage{latexsym}
\usepackage{amsmath}
\usepackage{amssymb}
\usepackage{amstext}
\usepackage{url}
\usepackage[noend]{algorithmic}
\usepackage{algorithm}
\usepackage{multirow}
\usepackage{bm}
\usepackage{enumitem}
\usepackage{microtype}
\usepackage{epstopdf}

\newtheorem{prop}{Proposition}
\newtheorem{defn}{Definition}

\newtheorem{ex}{Example}

\newcommand{\Real}{\mathbb R}

\newcommand{\aaa}{\mathbf{a}}
\newcommand{\m}{\mathbf{m}}
\newcommand{\s}{\mathbf{s}}
\newcommand{\C}{\mathcal{C}}
\newcommand{\T}{\mathcal{T}}
\newcommand{\OO}{\mathcal{O}}
\newcommand{\R}{\mathcal{R}}
\newcommand{\pomdp}{\textsf{POMDP}}
\newcommand{\ipomdp}{\textsf{I-POMDP}}

\begin{document}


\title{Individual Planning in Agent Populations: Exploiting Anonymity
  and Frame-Action Hypergraphs}

\author{Ekhlas Sonu, Yingke Chen \and Prashant Doshi \\
THINC lab, Dept. of Computer Science\\
University of Georgia\\
Athens, Georgia 30602\\
\{esonu,ykchen\}@uga.edu, pdoshi@cs.uga.edu}

\maketitle
\begin{abstract}
  Interactive   partially   observable   Markov   decision   processes
  (\ipomdp{})  provide   a  formal   framework  for  planning   for  a
  self-interested agent in multiagent settings.  An agent operating in
  a  multiagent environment  must  deliberate about  the actions  that
  other  agents may  take and  the effect  these actions  have  on the
  environment  and the  rewards it  receives.   Traditional \ipomdp{}s
  model this  dependence on  the actions of  other agents  using joint
  action and  model spaces.  Therefore, the  solution complexity grows
  exponentially  with  the   number  of  agents  thereby  complicating
  scalability.  In this paper, we model and extend {\em anonymity} and
  {\em  context-specific  independence}  -- problem  structures  often
  present  in  agent  populations   --  for  computational  gain.   We
  empirically demonstrate the efficiency from exploiting these problem
  structures by  solving a new multiagent problem  involving more than
  1,000 agents.
\end{abstract}

\section{Introduction}
\label{sec:intro}

We  focus  on  the  decision-making  problem of  an  individual  agent
operating  in  the  presence  of other  self-interested  agents  whose
actions  may affect  the  state  of the  environment  and the  subject
agent's rewards.  In stochastic and partially observable environments,
this  problem   is  formalized   by  the  {\em   interactive  \pomdp{}
  (\ipomdp{})}   ~\cite{Gmytrasiewicz05:Framework:JAIR}.    \ipomdp{}s
cover  an  important  portion   of  the  multiagent  planning  problem
space~\cite{Seuken07:Formal,Doshi12:Decision},   and  applications  in
diverse  areas  such as  security~\cite{Ng10:Towards,Seymour09:Trust},
robotics~\cite{Wang13:I-POMDP,Woodward12:Learning},       ad       hoc
teams~\cite{Chandrasekaran14:Team}       and       human      behavior
modeling~\cite{Doshi10:Modeling,Wunder11:Using}  testify  to its  wide
appeal while critically motivating better scalability.

Previous  \ipomdp{}   solution  approximations  such   as  interactive
particle     filtering~\cite{Doshi09:Monte},     point-based     value
iteration~\cite{Doshi08:Generalized}  and  interactive bounded  policy
iteration (I-BPI)~\cite{Sonu14:Scalable}  scale \ipomdp{} solutions to
larger  physical  state,  observation  and model  spaces.   Hoang  and
Low~(\citeyear{Hoang13:Interactive})    introduced   the   specialized
\ipomdp{} {\sf  Lite} framework  that promotes efficiency  by modeling
other  agents as  nested  {\sf MDP}s.   However,  to the  best of  our
knowledge no effort specifically scales \ipomdp{}s to many interacting
agents -- say,  {\em a population of more than  a thousand} -- sharing
the environment.

For illustration,  consider the decision-making problem  of the police
when faced with a large protest.  The degree of the police response is
often decided by how many protestors of which type (disruptive or not)
are  participating. The  individual identity  of the  protestor within
each type  seldom matters.  This key observation  of {\em frame-action
  anonymity} motivates us in how  we model the agent population in the
planning  process. Furthermore, the  planned degree  of response  at a
protest site is influenced, in part, by how many disruptive protestors
are predicted  to converge  at the  site and much  less by  some other
actions of  protestors such as  movement between other  distant sites.
Therefore, police  actions depend  on just a  few actions of  note for
each type of agent.

The example above illustrates two  known and powerful types of problem
structure  in domains involving  many agents:  \emph{action anonymity}
\cite{Roughgarden02:How}        and        {\em       context-specific
  independence}~\cite{Boutilier96:Context}.   Action  anonymity allows
the exponentially  large joint action  space to be substituted  with a
much  more  compact  space  of action  \emph{configurations}  where  a
configuration is a tuple  representing the number of agents performing
each action.   Context-specific independence (wherein  given a context
such as the state and agent's own action, not all actions performed by
other agents are  relevant) permits the space of  configurations to be
compressed by projecting counts over a limited set of others' actions.
We extend  both action anonymity and  context-specific independence to
allow    considerations    of    an    agent's    {\em    frame}    as
well.~\footnote{{\small  \ipomdp{}s  distinguish  between  an  agent's
    frame and type with the  latter including beliefs as well.  Frames
    are similar  in semantics  to the colloquial  use of  types.}}  We
list the specific contributions of this paper below:
  \begin{enumerate}[leftmargin=*,topsep=0in,itemsep=0in] 
  \item \ipomdp{}s are severely  challenged by large numbers of agents
    sharing the environment, which  cause an exponential growth in the
    space of  joint models and actions.   Exploiting problem structure
    in  the  form   of  frame-action  anonymity  and  context-specific
    independence, we present a new method for considerably scaling the
    solution of \ipomdp{}s to an unprecedented number of agents.
  \item We present a systematic  way of modeling the problem structure
    in transition,  observation and reward  functions, and integrating
    it in  a simple  method for solving  \ipomdp{}s that  models other
    agents  using finite-state machines  and builds  reachability trees
    given an initial belief.
  \item We prove that the  Bellman equation modified to include action
    configurations and frame-action  independences continues to remain
    optimal given the \ipomdp{} with explicated problem structure.
  \item  Finally,  we theoretically  verify  the  improved savings  in
    computational time and memory, and empirically demonstrate it on a
    new problem of policing protest with over a thousand protestors.
  \end{enumerate}

  The above  problem structure allows us to  emphatically mitigate the
  curse  of dimensionality whose  acute impact  on \ipomdp{}s  is well
  known.   However, it  does not  lessen the  impact of  the  curse of
  history. In this  context, an additional step of  sparse sampling of
  observations   while  generating   the   reachability  tree   allows
  sophisticated  planning with  a  population of  1,000+ agents  using
  about six hours.

\section{Related Work}
\label{sec:related}
Building  on  graphical  games~\cite{Kearns01:Graphical}, {\em  action
  graph games  (AGG)}~\cite{Jiang11:Action} utilize problem structures
such  as   action  anonymity  and   context-specific  independence  to
concisely represent  single shot complete-information  games involving
multiple  agents and  to  scalably solve  for  Nash equilibrium.   The
independence is modeled using a  directed action graph whose nodes are
actions and an edge between two  nodes indicates that the reward of an
agent performing an action indicated  by one node is affected by other
agents performing  action of  the other node.   Lack of  edges between
nodes encodes  the context-specific independence where  the context is
the action.  Action anonymity is useful when the action sets of agents
overlap substantially.   Subsequently, the  vector of counts  over the
set  of  distinct  actions,  called a  \emph{configuration},  is  much
smaller than the space of action profiles.

We substantially  build on AGGs  in this paper by  extending anonymity
and  context-specific  independence   to  include  agent  frames,  and
generalizing  their  use to  a  partially  observable stochastic  game
solved using decision-theoretic  planning as formalized by \ipomdp{}s.
Indeed,  Bayesian  AGGs~\cite{Jiang10:Bayesian}  extend  the  original
formulation  to include  agent types.   These result  in type-specific
action sets with the benefit  that the action graph structure does not
change   although   the   number    of   nodes   grows   with   types:
$|\hat{\Theta}||A|$ nodes for  agents with $|\hat{\Theta}|$ types each
having same $|A|$ actions.   If two actions from different type-action
sets  share a  node, then  these actions  are interchangeable.   A key
difference in our representation is that we explicitly model frames in
the graphs due to which context-specific independence is modeled using
frame-action  {\em  hypergraphs}.   Benefits  are  that  we  naturally
maintain the distinction between  two similar actions but performed by
agents  of  different  frames,  and  we  add  less  additional  nodes:
$|\hat{\Theta}|+ |A|$.   However, a hypergraph is a  more complex data
structure   for   operation.    Temporal 
AGGs~\cite{Jiang09:Temporal}  extend AGGs to  a repeated  game setting
and  allow decisions  to condition  on chance  nodes. These  nodes may
represent the  action counts from previous step  (similar to observing
the  actions in  the previous  game).  Temporal  AGGs come  closest to
multiagent   influence  diagrams~\cite{Koller01:Multi-agent}  although
they can additionally model  the anonymity and independence structure.
Overall, \ipomdp{}s  with frame-action anonymity  and context-specific
independence  significantly augment  the combination  of  Bayesian and
temporal AGGs  by utilizing the  structures in a  partially observable
stochastic game setting with agent types.

Varakantham  et  al.~(\citeyear{Varakantham14:Decentralized}) building
on  previous work~\cite{Varakantham12:Decision} recently  introduced a
decentralized MDP that models a simple form of anonymous interactions:
rewards and  transition probabilities specific to  a state-action pair
are  affected  by the  number  of  other  agents regardless  of  their
identities.  The  interaction influence  is not further  detailed into
which actions  of other agents  are relevant (as in  action anonymity)
and thus  configurations and  hypergraphs are not  used.  Furthermore,
agent types are not  considered.  Finally, the interaction hypergraphs
in   networked-distributed   \pomdp{}s~\cite{Nair05:Networked}   model
complete reward independence between  agents -- analogous to graphical
games -- which differs from  the hypergraphs in this paper (and action
graphs) that model independence in reward (and transition, observation
probabilities) along a different dimension: actions.

\section{Background}
\label{sec:background}

Interactive \pomdp{}s  allow a self-interested  agent to  plan individually  in a
partially  observable stochastic  environment in the  presence  of other
agents of uncertain types.   We briefly review the \ipomdp{} framework
and  refer  the  reader  to~\cite{Gmytrasiewicz05:Framework:JAIR}  for
further details.

A finitely-nested interactive  \ipomdp{} for an  agent (say  agent 0)
of strategy level $l$ operating  in a  setting inhabited  by one  of more  other interacting
agents is defined as the following tuple:
$$\text{\ipomdp{}}_{0,l} = \langle IS_{0,l}, A, T_0, \Omega_0,  O_0, R_0, OC_0 \rangle$$
\begin{itemize}[leftmargin=*,topsep=0in,itemsep=0in]
\item $IS_{0,l}$  denotes the set of {\em  interactive states} defined
  as, $IS_{0,l} = S  \times \prod_{j=1}^N M_{j,l-1}$, where $M_{j,l-1}
  = \{\Theta_{j,l-1} \cup SM_j\}$, for $l \geq 1$, and $IS_{i,0} = S$,
  where $S$ is the set of physical states. $\Theta_{j,l-1}$ is the set
  of   computable,   intentional  models   ascribed   to  agent   $j$:
  $\theta_{j,l-1}$ $=$  $\langle b_{j,l-1}, \hat{\theta}_j  \rangle $,
  where  $b_{j,l-1}$ is  agent $j$'s  level $l-1$  belief, $b_{j,l-1}$
  $\in$       $\triangle(IS_{j,l-1})$,       and      $\hat{\theta}_j$
  $\overset{\triangle}{=}$ $\langle A,  T_j, \Omega_j, O_j, R_j, OC_j
  \rangle$,   is  $j$'s   frame.    Here,  $j$   is   assumed  to   be
  Bayes-rational.   At {\em level  0}, $b_{j,0}$  $\in$ $\triangle(S)$
  and a level-0  intentional model reduces to a  POMDP.  $SM_j$ is the
  set of  subintentional models of $j$,  an example is  a finite state
  automaton;
\item $A$ $=$ $A_0 \times A_1  \times \ldots \times A_N$ is the set of
  joint actions of all agents;
\item $T_0: S \times A_0 \times \prod_{j=1}^N A_j \times S \rightarrow
  [0,1]$ is the transition function;
\item $\Omega_0$ is the set of agent 0's observations;
\item  $O_0: S  \times A_0  \times \prod_{j=1}^N  A_j  \times \Omega_0
  \rightarrow [0,1]$ is the observation
  function; 
\item $R_0:  S \times A_0 \times \prod_{j=1}^N  A_j \rightarrow \Real$
  is the reward function; and
\item $OC_0$ is  the optimality criterion, which is  identical to that
  for POMDPs.  In this paper, we consider  a finite-horizon optimality
  criteria.
\end{itemize}

Besides the  physical state  space, the \ipomdp{}'s  interactive state
space  contains all  possible models  of other  agents. In  its belief
update,  an agent has  to update  its belief  about the  other agents'
models based on an estimation about the other agents' observations and
how  they update their  models. As  the number  of agents  sharing the
environment grows, the size of the joint action and joint model spaces
increases  exponentially.   Therefore,   the  memory  requirement  for
representing  the transition, observation  and reward  functions grows
exponentially as  well as the  complexity of performing  belief update
over  the   interactive  states.   In  the  context   of  $N$  agents,
interactive bounded  policy iteration~\cite{Sonu14:Scalable} generates
good quality  solutions for an  agent interacting with 4  other agents
(total of 5 agents) absent any  problem structure.  To the best of our
knowledge, this result illustrates the best scalability so far to $N >
2$ agents.

\section{Many-Agent I-POMDP}
\label{sec:framework}

To  facilitate  understanding  and  experimentation,  we  introduce  a
pragmatic running example that also forms our evaluation domain.

\begin{figure}[ht]
  \centering
  \includegraphics[width=1.5in]{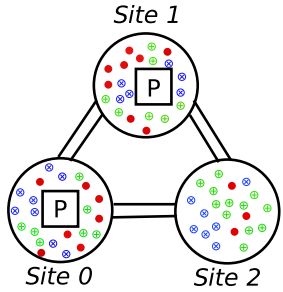}
  \caption{{\small Protestors of  different frames (colors) and police
      troops at  two of  three sites in  the policing  protest domain.
      The state space  of police decision making is  factored into the
      protest intensity levels at the sites.}}
  \label{fig:protest}
\end{figure}

\begin{ex} [Policing Protest]
  Consider a policing scenario  where police (agent $0$) must maintain
  order in  3 geographically distributed and  designated protest sites
  (labeled  0,  1,  and  2)  as shown  in  Fig.~\ref{fig:protest}.   A
  population of $N$ agents are  protesting at these sites.  Police may
  dispatch  one or  two  riot-control  troops to  either  the same  or
  different  locations.   Protests  with differing  intensities,  low,
  medium and high (disruptive), occur at each of the three sites.  The
  goal of the police is to deescalate protests to the low intensity at
  each  site.  Protest  intensity at  any  site is  influenced by  the
  number  of  protestors and  the  number  of  police troops  at  that
  location.  In the absence of  adequate policing, we presume that the
  protest intensity  escalates.  On the other hand,  two police troops
  at a location are adequate for de-escalating protests.
\label{Eg:MobControl}
\end{ex}

\subsection{Factored Beliefs and Update}
\label{subsec:belief}

As  we  mentioned  previously,  the  subject  agent  in  an  \ipomdp{}
maintains a belief  over the physical state and  joint models of other
agents, $b_{0,l} \in  \Delta(S \times \prod_{j=1}^N M_{j,l-1})$, where
$\Delta(\cdot)$  is  the  space  of  probability  distributions.   For
settings such  as Example~\ref{Eg:MobControl} where $N$  is large, the
size  of   the  interactive  state  space   is  exponentially  larger,
$|IS_{0,l}|  =   |S||M_{j,l-1}|^N$,  and  the   belief  representation
unwieldy.   However, the representation  becomes manageable  for large
$N$ if the belief is factored:
\begin{align}
&b_{0,l}(s,  m_{1,l-1}, m_{2,l-1},
\ldots, m_{N,l-1}) =  Pr(s)~Pr(m_{1,l-1}|s)\nonumber\\
& \times Pr(m_{2,l-1}|s) \times \ldots \times
Pr(m_{N,l-1}|s) 
\label{eqn:prior}
\end{align}
This  factorization  assumes  conditional  independence of  models  of
different agents given the  physical state. Consequently, beliefs that
correlate agents may not  be directly represented although correlation
could  be   alternately  supported   by  introducing  models   with  a
correlating device.

The memory consumed in storing a factored belief is $\mathcal{O}(|S| +
N|S||M_j^*|)$, where $|M_j^*|$ is the  size of the largest model space
among all other agents. This is  linear in the number of agents, which
is  much  less  than  the  exponentially growing  memory  required  to
represent  the belief  as a  joint distribution  over  the interactive
state space, $\mathcal{O}(|S||M_j^*|^N)$.

Given agent  0's belief at  time $t$, $b_{0,l}^t$, its  action $a_0^t$
and the subsequent observation it makes, $\omega_0^{t+1}$, the updated
belief at time step $t+1$, $b_{0,l}^{t+1}$, may be obtained as:
\begin{align}
&  Pr(s^{t+1}, m_{1,l-1}^{t+1}, \ldots, 
 m_{N,l-1}^{t+1}|b_{0,l}^t, a_0^t, \omega_0^{t+1})
= Pr(s^{t+1}| \nonumber\\
& b_{0,l}^t, a_0^t, \omega_0^{t+1}) 
~Pr(m_{1,l-1}^{t+1}|  
 s^{t+1},m_{2,l-1}^{t+1},\ldots, m_{N,l-1}^{t+1}, \nonumber\\
& b_{0,l}^t, a_0^t,
\omega_0^{t+1}) \times 
 \ldots \times 
 Pr(m_{N,l-1}^{t+1}|s^{t+1}, b_{0,l}^t, a_0^t, \omega_0^{t+1})
\label{eqn:BU_common}
\end{align}

Each factor in the  product of Eq.~\ref{eqn:BU_common} may be obtained
as follows. The update over the physical state
is: 
\begin{align}
& Pr(s^{t+1}| b_{0,l}^t, a_0^t, \omega_0^{t+1}) \propto Pr(s^{t+1}, \omega_0^{t+1} | b_{0,l}^t, a_0^t)\nonumber\\
& = \sum \limits_{s^t} b_{0,l}^t(s^t)  \sum \limits_{\m_{-0}^t}
b_{0,l}^t(m_{1,l-1}^t|s^t) \times \ldots \times b_{0,l}^t(m_{N,l-1}^t|s^t) \nonumber\\
&  \times \sum \limits_{\aaa_{-0}^t} Pr(a_1^t|m_{1,l-1}^t) \times \ldots \times Pr(a_N^t|m_{N,l-1}^t) \nonumber\\
& \times O_0^{t+1}(s^{t+1}, \langle a_0^t, \aaa_{-0}^t \rangle, \omega_0^{t+1})~T_0(s^t, \langle a_0^t, \aaa_{-0}^t \rangle, s^{t+1}) 
\label{eqn:update_state}
\end{align}
and the update over the model of each other agent, $j= 1 \ldots N$,
 conditioned on the state at $t+1$ is:
\begin{align}
  & Pr(m_{j,l-1}^{t+1}|s^{t+1}, m_{j+1,l-1}^{t+1}, \ldots, m_{N,l-1}^{t+1}, b_{0,l}^t, a_0^t, \omega_{0}^{t+1}) = \nonumber\\
  & \sum_{s^t} b_0^t(s^t) \sum_{\m_{-j,l-1}^t} b_{0,l}^t(m_{1,l-1}^t|s^t) \times \ldots \times b_{0,l}^t(m_{N,l-1}^t|s^t) \nonumber\\
  &  \sum_{\aaa_{-j}^t}
  Pr(a_1^t|m_{1,l-1}^t) \times \ldots \times Pr(a_n^t|m_{N,l-1}^t) \nonumber\\
  &    \sum_{\omega_j^{t+1}}   O_j(s^{t+1},\langle   a_j,   \aaa_{-j}^t
  \rangle, \omega_j^{t+1})~Pr(m_j^{t+1}|m_j^t, a_j^t, \omega_j^{t+1})
\label{eqn:update_model}
\end{align}
Derivations  of Eqs.~\ref{eqn:update_state} and~\ref{eqn:update_model}
are  straightforward and  not given  here due  to lack  of  space.  In
particular, note that models of agents  other than $j$ at $t+1$ do not
impact  $j$'s model  update  in the  absence  of correlated  behavior.
Thus,   under   the   assumption   of   a   factored   prior   as   in
Eq.~\ref{eqn:prior} and  absence of agent  correlations, the \ipomdp{}
belief update may  be decomposed into an update  of the physical state
and update of the models of $N$ agents conditioned on the state.

\subsection{Frame-Action Anonymity}
\label{subsec:anonymity}

As   noted   by   Jiang   et   al.~(\citeyear{Jiang11:Action}),   many
noncooperative  and cooperative  problems exhibit  the  structure that
rewards  depend on  the number  of  agents acting  in particular  ways
rather than which  agent is performing the act.   This is particularly
evident in  Example~\ref{Eg:MobControl} where the  outcome of policing
largely depends on the number  of protestors that are peaceful and the
number that are disruptive.  Building on this, we additionally observe
that  the transient  state of  the  protests and  observations of  the
police at a site are also largely influenced by the number of peaceful
and disruptive  protestors moving from one location  to another.  This
is noted in the example below:
\begin{ex} [Frame-action anonymity of protestors]
  The transient state of protests reflecting the intensity of protests
  at each  site depends on  the previous intensity  at a site  and the
  {\em  number} of  peaceful  and disruptive  protestors entering  the
  site.  Police  (noisily) observes the  intensity of protest  at each
  site which is again largely determined by the number of peaceful and
  disruptive protestors  at a site.  Finally, the  outcome of policing
  at a site  is contingent on whether the  protest is largely peaceful
  or  disruptive.   Consequently,   the  identity  of  the  individual
  protestors beyond their frame and action is disregarded.
\label{Eg:MobControlAnon}
\end{ex}
Here, peaceful and disruptive are  different {\em frames} of others in
agent 0's \ipomdp{},  and the above definition may  be extended to any
number of frames. Frame-action  anonymity is an important attribute of
the above  domain. We formally define  it in the context  of agent 0's
transition, observation and reward functions next:
\begin{defn} [Frame-action anonymity]
  Let $\aaa_{-0}^p$ be  a joint action of all  peaceful protestors and
  $\aaa_{-0}^d$  be  a  joint  action  of all  disruptive  ones.   Let
  $\dot{\aaa}_{-0}^p$ and  $\dot{\aaa}_{-0}^d$ be permutations  of the
  two  joint   action  profiles,  respectively.    An  I-POMDP  models
  frame-action anonymity  iff for any $a_0$,  $s$, $s'$, $\aaa_{-0}^p$
  and $\aaa_{-0}^d$:\\
  $T_0(s$, $a_0, \aaa_{-0}^p, \aaa_{-0}^d, s') =
  T_0(s, a_0, \dot{\aaa}_{-0}^p, \dot{\aaa}_{-0}^d, s')$,\\
  $O_0(s', a_0, \aaa_{-0}^p, \aaa_{-0}^d, \omega_0) = O_0(s',
  a_0, \dot{\aaa}_{-0}^p, \dot{\aaa}_{-0}^d, \omega_0)$,  and \\
  $R_0(s,   a_0,   \aaa_{-0}^p,    \aaa_{-0}^d   )   =   R_0(s,   a_0,
  \dot{\aaa}_{-0}^p, \dot{\aaa}_{-0}^d)$ ~~$\forall$
  $\dot{\aaa}_{-0}^p$, $\dot{\aaa}_{-0}^d$.
\label{def:anon}
\end{defn}
Recall the definition of an  action configuration, $\C$, as the vector
of  action counts  of an  agent  population.  A  permutation of  joint
actions of others,  say $\dot{\aaa}_{-0}^p$, assigns different actions
to individual agents.  Despite this,  the fact that the transition and
observation probabilities, and  the reward remains unchanged indicates
that the  identity of the  agent performing the action  is irrelevant.
Importantly, the configuration of the joint action and its permutation
stays  the  same:  $\C(\aaa_{-0}^p)  =  \C(\dot{\aaa}_{-0}^p)$.   This
combined  with Def.~\ref{def:anon}  allows redefining  the transition,
observation  and  reward  functions  to  be  over  configurations  as:
$T_0(s,a_0,\C(\aaa_{-0}^p),\C(\aaa_{-0}^d),       s')$,       $O_0(s',
a_0,\C(\aaa_{-0}^p),\C(\aaa_{-0}^d),       o)$       and       $R_0(s,
a_0,\C(\aaa_{-0}^p),\C(\aaa_{-0}^d$ $))$.

Let $A_1^p$,  \ldots, $A_n^p$  be overlapping sets  of actions  of $n$
peaceful protestors, and $A_{-0}^p$  is the Cartesian product of these
sets.  Let $\C(A_{-0}^p)$ be the  set of all action configurations for
$A_{-0}^p$. {\em  Observe that multiple joint  actions from $A_{-0}^p$
  may  result  in a  single  configuration;  these  joint actions  are
  configuration equivalent.}  Consequently, the equivalence partitions
the  joint  action   set  $A_{-0}^p$  into  $|\C(A_{-0}^p)|$  classes.
Furthermore, when other agents of  same frame have overlapping sets of
actions, the number  of configurations could be much  smaller than the
number  of joint  actions. Therefore,  definitions of  the transition,
observation  and reward  functions involving  configurations  could be
more compact.

\subsection{Frame-Action Hypergraphs}
\label{subsec:hypergraphs}

In  addition  to   frame-action  anonymity,  domains  involving  agent
populations  often exhibit context-specific  independences. This  is a
broad category and includes the context-specific independence found in
conditional        probability        tables        of        Bayesian
networks~\cite{Boutilier96:Context}  and  in  action-graph games.   It
offers    significant   additional    structure    for   computational
tractability.   We  begin  by  illustrating  this in  the  context  of
Example~\ref{Eg:MobControl}.

\begin{ex}[Context-specific  independence   in  policing]  At   a  low
  intensity protest site, reward for the police on passive policing is
  independent of  the movement  of the protestors  to other  sites.
  The transient intensity of the protest  at a site given the level of
  policing at the site (context)
  is independent of the movement of protestors between other sites.
\end{ex} 

The  context-specific   independence  above  builds   on  the  similar
independence in action graphs in two ways: $(i)$ We model such partial
independence  in  the  transitions  of  factored  states  and  in  the
observation  function as  well, in  addition to  the  reward function.
$(ii)$ We  allow the context-specific  independence to be  mediated by
the frames of other agents in addition to their actions.  For example,
the rewards received from policing a site is independent of the number
of protestors at  another site, instead the rewards  are influenced by
the number  of {\em peaceful} and {\em  disruptive} protestors present
at that site.

\begin{figure}[ht]
  \centering
  \includegraphics[width=1.625in]{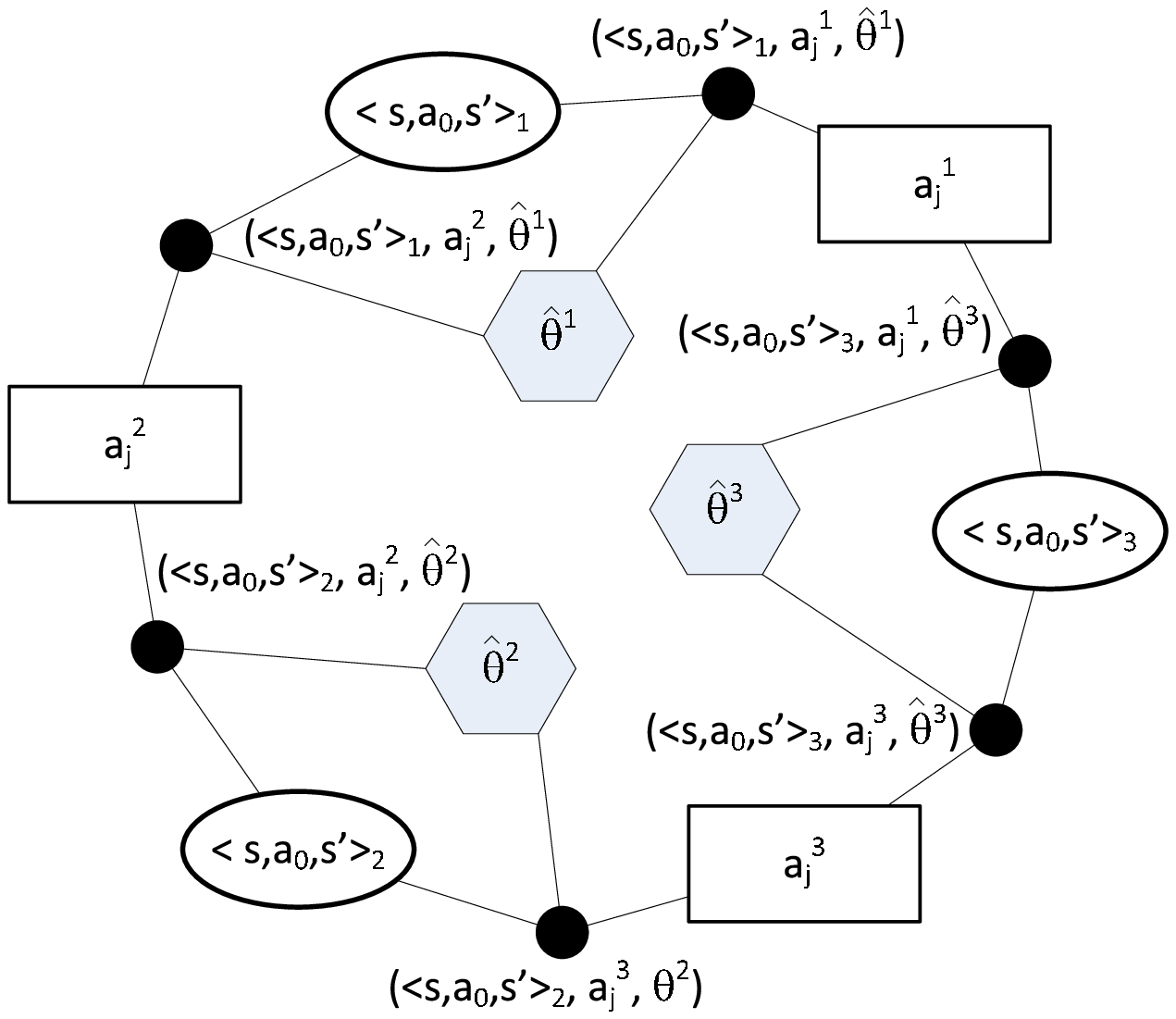}
  \includegraphics[width=1.625in]{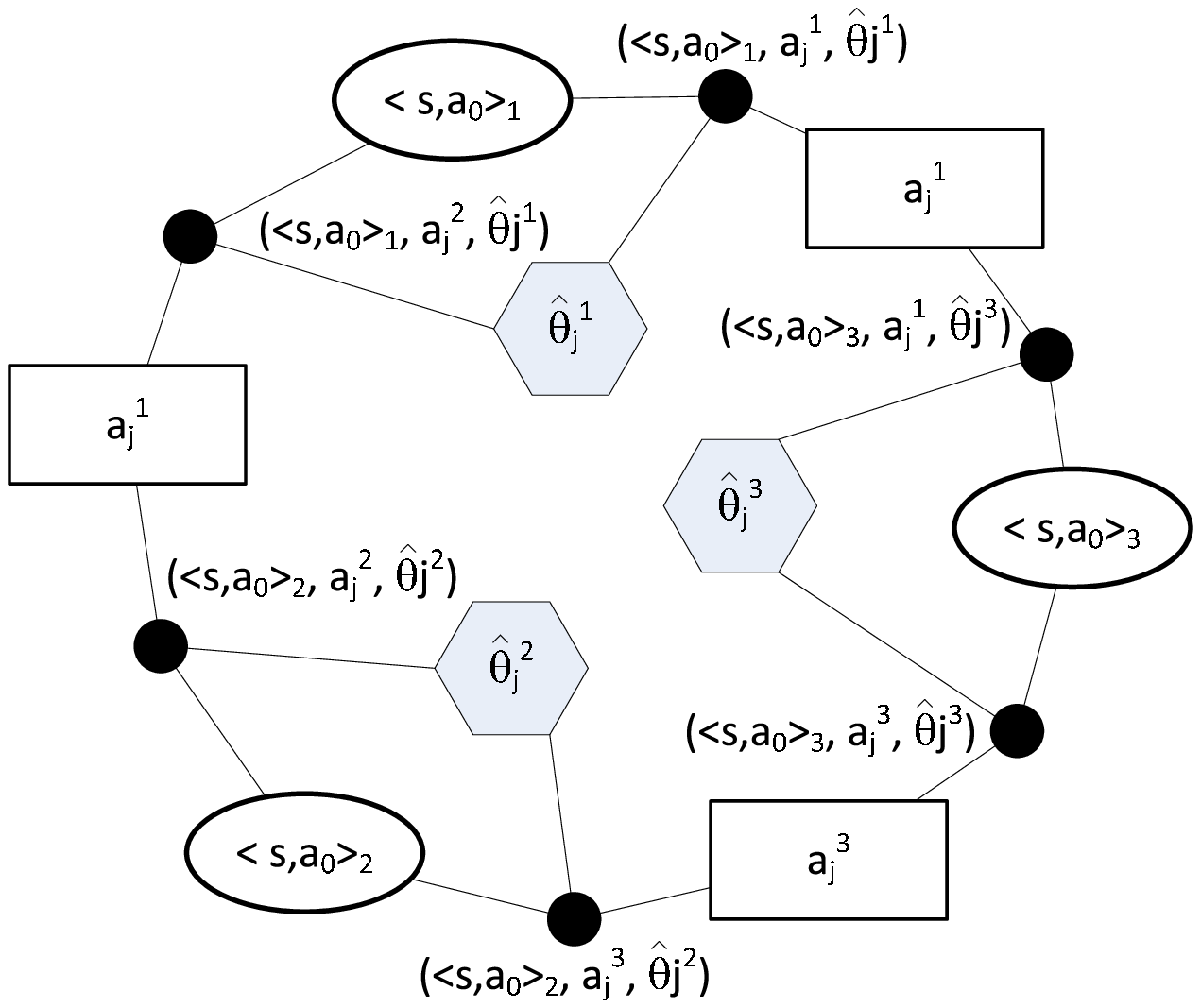}
  {\small $(a)$ \hspace{1.65in} $(b)$}
  \caption{{\small Levi (incidence)  graph representation of a generic
      frame-action hypergraph  for $(a)$ the  transition function, and
      $(b)$ the reward function.   The shaded nodes represent edges in
      the hypergraph.  Each edge has  the context, $\psi$,  denoted in
      bold,  agent's  action,  $a$,  and  its  frame,  $\hat{\theta}$,
      incident on  it. For example, the  reward for a  state and agent
      0's action,  $\langle s, a_0  \rangle_1$ is affected  by others'
      actions  $a_j^1$ and  $a_j^2$ performed  by any  other  agent of
      frame $\hat{\theta}_j^1$ only.}}
  \label{fig:Levi}
\end{figure}

The  latter  difference  generalizes   the  action  graphs  into  {\em
  frame-action  hypergraphs}, and  specifically  3-uniform hypergraphs
where  each  edge is  a  set  of 3  nodes.   We  formally define
it below:

\begin{defn} [Frame-action hypergraph]
  A frame-action  hypergraph for agent  $0$ is a  3-uniform hypergraph
  $\mathcal{G} = \langle \Psi, A_{-0}, \hat{\Theta}_{-0}, E \rangle$, where
  $\Psi$ is a set of nodes that represent the context, $A_{-0}$ is a set of
  action nodes with each node representing an action that any other agent may take; $\hat{\Theta}_{-0}$ is a set of frame nodes, each
  node  representing  a frame  ascribed  to an  agent,  and  $E$ is  a
  3-uniform hyperedge  containing one node from each  set $\Psi$, $A_{-0}$,
  and $\hat{\Theta}_{-0}$, respectively.
\label{Def:TAG}
\end{defn}

Both context and  action nodes differ based on  whether the hypergraph
applies to  the transition, observation or reward  functions:
\begin{itemize}[leftmargin=*,topsep=0in,itemsep=0in]
\item For the transition function, the context is the set of all pairs
  of states  between which a transition  may occur and  each action of
  agent $0$,  $\Psi =  S \times  A_0 \times S$,  and the  action nodes
  includes actions of all  other agents, $A_{-0}$ $=$ $\bigcup_{j=1}^N
  A_j$.  Neighbors  of a context node  $\psi$ $=$ $\langle  s, a_0, s'
  \rangle$ are all the  frame-action pairs that affect the probability
  of  the  transition.   An   edge  $(\langle$  $s,  a_0,  s'\rangle$,
  $a_{-0}$,  $\hat{\theta})$  indicates  that  the  probability  of
  transitioning from $s$  to $s'$ on performing $a_0$  is affected (in
  part)  by the other  agents of  frame $\hat{\theta}$  performing the
  particular action in $A_{-0}$.
\item  The  context  for  agent  $0$'s  observation  function  is  the
  state-action-observation  triplet,  $\Psi  =  S  \times  A_0  \times
  \Omega_0$,  and the  action  nodes  are identical  to  those in  the
  transition function.  Neighbors of a context node,  $\langle s, a_0,
  \omega_0 \rangle$, are all  those frame-action pairs that affect the
  observation  probability. Specifically,  an edge  $(\langle  s, a_0,
  \omega_0  \rangle,  a_{-0},  \hat{\theta})$  indicates  that  the
  probability  of observing  $\omega_0$ from  state $s$  on performing
  $a_0$ is affected  (in part) by the other  agents performing action,
  $a_{-0}$, who possess frame $\hat{\theta}$.
\item For agent  0's reward function, the context is  the set of pairs
  of  state and action  of agent  0, $\Psi  = S  \times A_0$,  and the
  action  nodes  the  same  as  those in  transition  and  observation
  functions.  
  An   edge  $(\langle$  $s,a_0  \rangle$,
  $a_{-0}$, $\hat{\theta}_{-0})$ in  this hypergraph indicates that
  the reward  for agent 0 on  performing action $a_0$ at  state $s$ is
  affected (in  part) by the  agents of frame  $\hat{\theta}_{-0}$ who
  perform action in $A_{-0}$.
\end{itemize}
We illustrate  a general frame-action  hypergraph for context-specific
independence  in  a  transition  function  and a  reward  function  as
bipartite   Levi  graphs   in  Figs.~\ref{fig:Levi}$(a)$   and  $(b)$,
respectively. We point out that the hypergraph for the reward function
comes   closest  in   semantics   to  the   graph   in  action   graph
games~\cite{Jiang11:Action} although the former  adds the state to the
context and  frames.  Hypergraphs  for the transition  and observation
functions  differ  substantially in  semantics  and  form from  action
graphs.

To  use these  hypergraphs  in  our algorithms,  we  first define  the
general \emph{frame-action neighborhood} of a context node.

\begin{defn} [Frame-action neighborhood]
  The  frame-action neighborhood of  a context  node $\psi  \in \Psi$,
  $\nu(\psi)$,  given  a   frame-action  hypergraph  $\mathcal{G}$  is
  defined as a subset of  $A \times \hat{\Theta}$ such that $\nu(\psi)
  =  \{(a_{-0}, \hat{\theta})|  a_{-0} \in  A_{-0}, \hat{\theta}
  \in \hat{\Theta}, (\psi, a_{-0}, \hat{\theta}) \in E\}$.
\end{defn}
As an  example, the frame-action neighborhood of  a state-action pair,
$\langle s,a_0 \rangle$ in a hypergraph for the reward function is the
set of all action and  frame nodes incident on each hyperedge anchored
by the node $\langle s,a_0 \rangle$.

We  move  toward  integrating  frame-action  anonymity  introduced  in
the previous subsection with  the context-specific independence
as modeled above by introducing frame-action configurations. 

\begin{defn} [Frame-action configuration]
  A  configuration over  the  frame-action neighborhood  of a  context
  node,  $\psi$,  given a  frame-action  hypergraph  is  a vector,  $$
  \C^{\nu(\psi)}        \overset{\triangle}{=}        \langle        ~
  \C(A^{\hat{\theta}_1}_{-0}),       \C(A^{\hat{\theta}_2}_{-0}),
  \ldots,  \C(A^{\hat{\theta}_{|\hat{\Theta}|}}_{-0}),  \C(\phi)  ~
  \rangle$$ where  each $a$ included  in $A^{\hat{\theta}}_{-0}$ is
  an   action   in   $\nu(\psi)$   with  frame   $\hat{\theta}$,   and
  $\mathcal{C}(A^{\hat{\theta}}_{-0})$  is   a  configuration  over
  actions by agents  other than 0 whose frame  is $\hat{\theta}$.  All
  agents with frames other than those in the frame-action neighborhood
  are assumed to perform a dummy action, $\phi$.
\label{def:Conf_TAG}
\end{defn}

Definition~\ref{def:Conf_TAG}  allows further inroads  into compacting
the  transition, observation  and rewards  functions of  the \ipomdp{}
using  context-specific independence.   Specifically, we  may redefine
these functions  one more time  to limit the configurations  only over
the frame-action neighborhood of the context as, $T_0(s,a_0,\C^{\nu(s,
  a_0, s')},  s')$, $O_0(s',a_0,\C^{\nu(s', a_0, \omega_0)},\omega_0)$
and $R_0(s, a_0,\C^{\nu(s,a_0)})$.~\footnote{Context in our transition
  function is  $\langle$$s,a_0,s'$$\rangle$ compared with  the context
  of     just    $\langle$$s,a_0$$\rangle$    in     Varakantham    et
  al's~(\citeyear{Varakantham14:Decentralized}) transitions.}

\subsection{Revised Framework}
\label{subsec:ipomdp_manyagents}

To benefit  from structures of anonymity and context-specific
independence, we redefine \ipomdp{} for agent $0$ as:
$$\text{\ipomdp{}}_{0,l} = \langle IS_{0,l}, A, \Omega_0, \T_0, \OO_0, \R_0, OC_0 \rangle$$
where:
\begin{itemize}[leftmargin=*,topsep=0in,itemsep=0in]
\item $IS_{0,l}$,  $A$, $\Omega_0$ and  $OC_0$ remain  the same  as
  before.  The  physical states are  factored as, $S  = \prod_{k=1}^K
  X_k$.
\item $\T_0$ is the transition function, $\T_0(x, a_0, \C^{\nu(x, a_0,
    x')},x')$  where $\C^{\nu(x, a_0,x')}$  is the  configuration over
  the  frame-action  neighborhood  of  context  $\langle  x,  a_0,  x'
  \rangle$ obtained  from a hypergraph  that holds for  the transition
  function.  This transition  function is  significantly  more compact
  than       the        original       that       occupies       space
  $\mathcal{O}(|X|^2|A_0||A_{-0}|^{N})$      compared      to      the
  $\mathcal{O}(|X|^2|A_0|(\frac{N}{|\nu^*|})^{|\nu^*|})$   of  $\T_0$,
  where      the      fraction      is     the      complexity      of
  $\binom{N+|\nu|^*+1}{|\nu^*|+1}$,    $|\nu^*|$   is    the   maximum
  cardinality    of   the   neighborhood    of   any    context,   and
  $(\frac{N}{|\nu^*|})^{|\nu^*|}$ $\ll$ $|A_{-0}|^N$. The value $\binom{N+|\nu|^*}{|\nu|^*}$ is obtained from combinatorial compositions and represents the number of ways $|\nu^*| + 1$ non-negative values can be weakly composed such that their sum is $N$. 
\item   The  redefined   observation  function   is   $\OO_0(x',  a_0,
  \C^{\nu(x',  a_0, \omega_0)}  , \omega_0)$  where  $\C^{\nu(x', a_0,
    \omega_0)}$   is   the   configuration   over   the   frame-action
  neighborhood of context $\langle x', a_0, \omega_0 \rangle$ obtained
  from  a   hypergraph  that  holds  for   the  observation  function.
  Analogously  to the  transition function,  the  original observation
  function  consumes  space $\mathcal{O}(|X||\Omega||A_0||A_{-0}|^N)$,
  which         is        much        larger         than        space
  $\mathcal{O}(|X||\Omega||A_0|(\frac{N}{|\nu^*|})^{|\nu^*|})$
  occupied by this redefinition.
\item  $\R_0$   is  the   reward  function  defined   as  $\R_0(x,a_0,
  \C^{\nu(x,a_0)})$ where $\C^{\nu(x,a_0)}$  is defined analogously to
  the  configurations in  the previous  parameters. The  reward  for a
  state and  actions may simply  be the sum  of rewards for  the state
  factors and actions (or a more general function if needed).  As with
  the transition  and observation  functions, this reward  function is
  compact                        occupying                       space
  $\mathcal{O}(|X||A_0|(\frac{N}{|\nu^*|})^{|\nu^*|})$  that  is  much
  less than $\mathcal{O}(|X||A_0||A_{-0}|^N)$ of the original.
\end{itemize}

\subsubsection{Belief Update} 
For  this extended  I-POMDP,  we  compute the  updated  belief over  a
physical    state    as   a    product    of    its   factors    using
Eq.~\ref{eqn:update_state_conf} and  belief update over  the models of
each other agent using Eq~\ref{eqn:update_model_conf} as shown below:
\begin{align}
& Pr(\s^{t+1}| b_{0,l}^t, a_0^t, \omega_0^{t+1}) \propto  \bigg \{ \sum \limits_{\s^t} b_{0,l}^t(\s^t) \prod \limits_{k=1}^{K}  \sum_{\C^{\nu(x_k^{t+1},a_0^t,\omega_{0}^{t+1})}} \nonumber\\
& Pr(\C^{\nu(x_k^{t+1},a_0^t,\omega_{0}^{t+1})}| b_{0,l}^t(M_{1,l-1}|\s^t), \ldots, b_{0,l}^t(M_{N,l-1}|\s^t)) \nonumber\\
& \OO_0(x_k^{t+1},a_0^t,\C^{\nu(x_k^{t+1}, a_0^t,\omega_{0}^{t+1})}, \omega_{0}^{t+1}) \bigg \}  \times \bigg \{ \sum \limits_{\s^t} b_{0,l}^t(\s^t) \prod \limits_{k=1}^{K} \nonumber \\
& \sum_{\C^{\nu(x_k^t,a_0^t,x_k^{t+1})}}
Pr(\C^{\nu(x_k^t,a_0^t,x_k^{t+1})}|b_{0,l}^t(M_{1,l-1}|\s^t), \ldots, \nonumber\\
&  b_{0,l}^t(M_{N,l-1}|\s^t))\T_0(x_k^t, \C^{\nu(x_k^t,a_0^t, x_k^{t+1})}, x_k^{t+1}) \bigg \}
\label{eqn:update_state_conf}
\end{align}
Here,  the  term,  $Pr(\C^{\nu(x_k^{t+1},  a_0^t,  \omega_{0}^{t+1})}|
b_{0,l}^t(M_{1,l-1}|s^t), \ldots,$ $b_{0,l}^t(M_{N,l-1}|s^t))$, is the
probability      of     a     frame-action      configuration     (see
Def.~\ref{def:Conf_TAG})  that  is context  specific  to the  triplet,
$\langle x^{t+1},a_0,\omega^{t+1}  \rangle$.  It is  computed from the
factored  beliefs over  the models  of  all others.   We discuss  this
computation in the next section.  The second configuration term has an
analogous meaning and is computed similarly.

The factored belief update over the  models of each other agent, $j= 1
\ldots N$, conditioned on the state at $t+1$ becomes:
\begin{align}
& Pr(m_{j,l-1}^{t+1}|s^{t+1}, \m_{-j,l-1}^{t+1}, b_{0,l}^t, a_0^t) =
\sum_{\s^t} b_0^t(\s^t) \sum_{m_j^t} b_0^t(m_j^t \nonumber\\ 
& |s^t) \sum_{a_j^t} Pr(a_j^t|m_j^t) \sum_{\C^{\nu(x^{t+1},a_j^t,\omega_{j})}}  Pr(\C^{\nu(x^{t+1},a_j^t,\omega_j)}| \nonumber\\
& b_{0,l}^t(M_{1,l-1}|\s^t),\ldots, b_{0,l}^t(M_{N,l-1}|\s^t)) \sum_{o_j^{t+1}} \OO_j(x^{t+1}, \nonumber\\
& a_j^t,\C^{\nu(x^{t+1},a_j^t,\omega_j)},\omega_j^{t+1})~Pr(m_j^{t+1}|m_j^t, a_j^t, \omega_j^{t+1}) 
\label{eqn:update_model_conf}
\end{align}

Proofs       for       obtaining      Eqs.~\ref{eqn:update_state_conf}
and~\ref{eqn:update_model_conf} are omitted due to space restrictions.
Notice that  the distributions over configurations  are computed using
distributions over  other agents' models. Therefore,  we must maintain
and update conditional beliefs  over other agents' models.  Hence, the
problem cannot  be reduced to  a \pomdp{} by  including configurations
with physical states.

\subsubsection{Value Function}
\label{subsec:value}

The  finite-horizon   value  function  of   the  many-agent  \ipomdp{}
continues  to be  the  sum of  agent  0's immediate  reward and  the
discounted expected reward over the future:
\begin{align}
V^h(m_{0,l}^t) =& \max_{a_0^t \in A_0} ~ER_0(b_{0,l}^t, a_0^t) + \nonumber\\
& \gamma \sum \limits_{\omega_0^{t+1}} Pr(\omega_0^{t+1}|b_{0,l}^t, a_0^t) V^{h-1}(m_{0,l}^{t+1})
\label{eqn:value}
\end{align}
where  $ER_0(b_{0,l}^t, a_0^t)$  is the  expected immediate  reward of
agent $0$ and  $\gamma$ is the discount factor. In  the context of the
redefined reward function of  the many-agent \ipomdp{} framework in this section,
the expected immediate reward is obtained as:
\begin{align}
& ER_0(b_{0,l}^t, a_0^t) =  \sum_{\s^t} b_{0,l}^t(\s^t) \bigg ( 
 \sum_{k=1}^K \sum_{\C^{\nu(x_k^t,a_0^t)}}
 Pr(\C^{\nu(x_k^t,a_0^t)}| \nonumber\\
& b_{0,l}^t(M_{1,l-1}|\s^t),  \ldots , b_{0,l}^t(M_{N,l-1}|\s^t)) R_0(x_k^t,a_0^t,
\C^{\nu(x_k^t,a_0^t})  \bigg )
\label{eqn:exp_reward}
\end{align}
where the  outermost sum is  over all the  state factors, $\s^t  = \langle
x_1^t,\ldots,x_K^t        \rangle$,        and        the        term,
$Pr(\C^{\nu(x_k^t,a_0^t)}|b_{0,l}^t(M_{1,l-1}|\s^t),      \ldots      ,
b_{0,l}^t(M_{N,l-1}|\s^t))$ denotes  the probability of  a frame-action
configuration that is context-specific to the factor, $x_k^t$. Importantly, Proposition~\ref{prop:optimality} establishes 
that the Bellman equation above is exact. The proof is given in the extended version of this paper~\cite{Sonu15:Individual:arXiv}. 

\begin{prop}[Optimality] The dynamic programming in Eq.~\ref{eqn:value} provides an exact computation 
of the value function for the many-agent \ipomdp{}.
\label{prop:optimality}
\end{prop}

\section{Algorithms}
\label{sec:algo}

We  present   an  algorithm  that   computes  the
distribution  over frame-action  configurations and  outline our
simple method for solving the many-agent \ipomdp{} defined previously.

\subsection{Distribution Over Frame-Action Configurations}
\label{subsec:algo_conf}

Algorithm~\ref{alg:conf}  generalizes   an  algorithm  by   Jiang  and
Lleyton-Brown~(\citeyear{Jiang11:Action})  for computing configurations over
actions given mixed  strategies of other agents to  include frames and
conditional  beliefs over  models of  other agents.   It  computes the
probability  distribution  of  configurations  over  the  frame-action
neighborhood of  an action given  the belief over the  agents' models:
$Pr(\C^{\nu(x,a_0,\omega_0)}|b_{0,l}^t(M_{1,l-1}|\s^t),$            $\ldots,
b_{0,l}^t(M_{N,l-1}|\s^t))$ and
$Pr(\C^{\nu(x,a_0,x')}|b_{0,l}^t(M_{1,l-1}|\s^t),       \ldots,$
$b_{0,l}^t(M_{N,l-1}|\s^t))$ in Eq.~\ref{eqn:update_state_conf},
$Pr(\C^{\nu(x,\omega_j)}|    b_{0,l}^t(M_{1,l-1}|\s^t),   \ldots,$
$b_{0,l}^t(M_{N,l-1}|\s^t))$\\ in Eq.~\ref{eqn:update_model_conf}, and
$Pr(\C^{\nu(x,a_0)}|b_{0,l}^t(M_{1,l-1}|\s^t),      \ldots    ,$
$b_{0,l}^t(M_{N,l-1}|\s^t))$ in Eq.~\ref{eqn:exp_reward}.

\begin{algorithm}
\caption{Computing $Pr(\C^{\nu(\cdot)}|b_{0,l}(M_{1,l-1}|\s),$ $\ldots , b_{0,l}(M_{N,l-1}|\s))$ }
\label{alg:DistConfig}
\textbf{Input:} $\nu(\cdot)$, $\langle b_{0,l}(M_{1,l-1}|\s), \ldots , b_{0,l}(M_{N,l-1}|\s) \rangle$  \\
\textbf{Output:}  A  trie  $\mathcal{P}_n$  representing distribution  over  the
frame-action configurations over $\nu(\cdot)$
\begin{algorithmic}[1]
  \STATE  Initialize $c_{0} \leftarrow (0, \ldots , 0)$, one value for each frame-action pair in $\nu(\cdot)$ and for $\phi$. Insert into empty trie $\mathcal{P}_0$

  \STATE Initialize $\mathcal{P}_0[c_{0}] \leftarrow 1$

  \FOR  {$j \leftarrow 1$ to  $N$}

  \STATE Initialize $\mathcal{P}_j$ to be an empty trie

  \FORALL {$c_{j-1}$ from $\mathcal{P}_{j-1}$}

  \FORALL {$m_{j,l-1} \in M_{j,l-1}$}

  \FORALL {$a_j \in A_j$ such that $Pr(a_j|m_{j,l-1}) > 0 $}

  \STATE  $c_{j} \leftarrow c_{j-1}$

  \IF {$\langle a_j,\hat{\theta}_j \rangle \in \nu(\cdot)$}
  \STATE  $c_{j}[a_j] \leftarrow c_{j}[a_j] + 1$  
  \ELSE
  \STATE  $c_{j}[\phi] \leftarrow c_{j}[\phi] + 1$
  \ENDIF

  \IF {$\mathcal{P}_j[c_{j}]$ does not exist }

  \STATE Initialize $\mathcal{P}_j[c_{j}] \leftarrow 0$

  \ENDIF

  \STATE $\mathcal{P}_j[c_{j}] \leftarrow \mathcal{P}_j[c_{j}] + \mathcal{P}_{j-1}[c_{j-1}] \times Pr(a_j|m_{j,l-1}) \times b_{0,l}(m_{j,l-1}|\s) $

  \ENDFOR

  \ENDFOR
  \ENDFOR

  \ENDFOR
  \RETURN {$\mathcal{P}_n$}
\end{algorithmic}
\label{alg:conf}
\vspace{-0.05in}
\end{algorithm}

Algorithm~\ref{alg:conf} adds the actions of each agent one at a time. A Trie data structure enables efficient insertion and access of the configurations. We begin by initializing the configuration space for 0 agents ($\mathcal{P}_0$) to contain one tuple of integers ($c_0$) with $|\nu|+1$ 0s and assign its probability to be 1 (lines 1-2). Using the configurations of the previous step, we construct the configurations over the actions performed by $j$ agents by adding 1 to a relevant element depending on $j$'s action and frame  (lines 3-15). If an action $a_j$ performed by $j$ with frame $\hat{m_j}$ is in the frame-action neighborhood $\nu(\cdot)$, then we increment its corresponding count by 1. Otherwise, it is considered as a dummy action and the count of $\phi$ is incremented (lines 9-12). Similarly, we update the probability of a configuration using the probability of $a_j$ and that of the base configuration $c_{j-1}$ (line 15). This algorithm is invoked multiple times for different values of $\nu(\cdot)$ as needed in the belief update and value function computation.

We utilize a simple method for solving the many-agent \ipomdp{} given an initial belief: each other agent is modeled using a finite-state controller as part of the interactive state space. A reachability tree of beliefs as nodes is projected for as many steps as the horizon (using Eqs.~\ref{eqn:update_state_conf} and~\ref{eqn:update_model_conf}) and value iteration (Eq.~\ref{eqn:value}) is performed on the tree. In order to mitigate the curse of history due to the branching factor that equals the number of agent 0's actions and observations, we utilize the well-known technique of sampling observations from the propagated belief and obtain a sampled tree on which value iteration is run to get a policy. Action for any observation that does not appear in the sample is that which maximizes the immediate expected reward.   

\section{Computational Savings}

The complexity of accessing an element in a ternary search trie is $\Theta(\nu)$. The maximum number of configurations encountered at any iteration is upper bounded by total number of configurations for $N$ agents, i.e. $\mathcal{O}((\frac{N}{|\nu^*|})^{|\nu^*|})$. The complexity of Algorithm~\ref{alg:conf} is {\em polynomial} in $N$,  $\mathcal{O}(N|M_j^*||A_j^*||\nu^*|(\frac{N}{|\nu^*|})^{|\nu^*|})$ where $M_j^*$ and $A_j^*$ are largest sets of models and actions for any agent.

For the traditional \ipomdp{} belief update, the complexity of computing Eq.~\ref{eqn:update_state} is $\mathcal{O}(|S||M_j^*|^{N}|A_j^*|^N)$ and that for computing Eq.~\ref{eqn:update_model} is $\mathcal{O}(|S||M_j^*|^N|A_j^*|^N|\Omega_j^*|)$ where $^*$ denotes the maximum cardinality of a set for any agent. For a factored representation, belief update operator invokes Eq. \ref{eqn:update_state} for each value of all state factors and it invokes Eq.~\ref{eqn:update_model} for each model of each agent $j$ and for all values of updated states. Hence the total complexity of belief update is $\mathcal{O}(N|M_j^*||S|^2|M_j^*|^N|A_j^*|^N|\Omega_j^*|)$. The complexity of computing updated belief over state factor $x^{t+1}$ using Eq.~\ref{eqn:update_state_conf} is $\mathcal{O}(|S|NK|M_j^*||A_j^*||\nu^*|(\frac{N}{|\nu^*|})^{|\nu^*|})$ (recall the complexity of Algorithm \ref{alg:conf}). Similarly, the complexity of computing updated model probability using Eq.~\ref{eqn:update_model_conf} is  $\mathcal{O}((|S|N|M_j^*||A_j^*||\nu^*|+|\Omega_j^*|)(\frac{N}{|\nu^*|})^{|\nu^*|})$. These complexity terms are polynomial in $N$ for small values of $|\nu^*|$ as opposed to exponential in $N$ as in Eqs.~\ref{eqn:update_state} and \ref{eqn:update_model}. The overall complexity of belief update is also polynomial in $N$. 

Complexity of computing the immediate expected reward in the absence of problem structure is $\mathcal{O}(|S|K|M_j^*|^N|A_j^*|^N)$. On the other hand, the complexity of computing expected reward using Eq.~\ref{eqn:exp_reward} is $\mathcal{O}(|S|KN|M_j^*||A_j^*||\nu^*|(\frac{N}{|\nu^*|})^{|\nu^*|})$, which is again polynomial in $N$ for low values of $|\nu^*|$. These complexities are discussed in greater detail in~\cite{Sonu15:Individual:arXiv}.

\section{Experiments}

\begin{figure}[ht]
  \centering
  \includegraphics[width=1.625in]{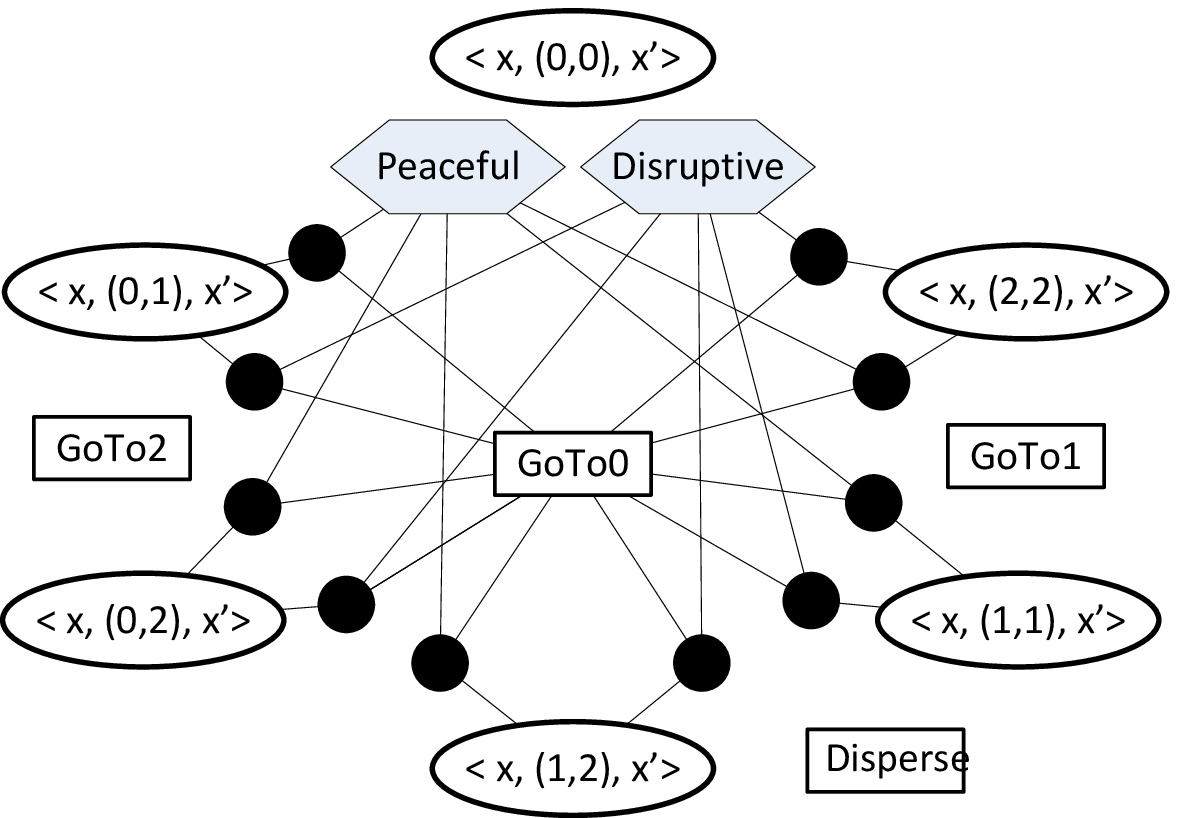}
  \includegraphics[width=1.625in]{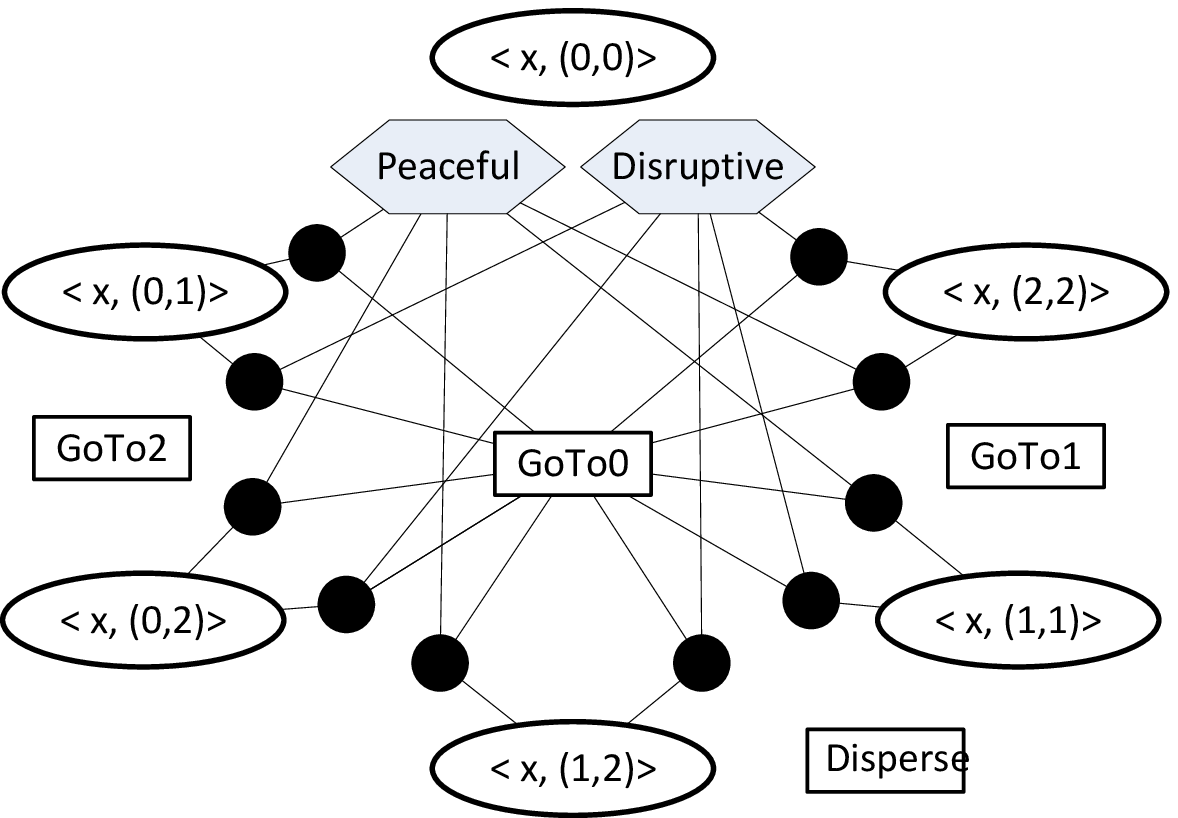}
  {\small $(a)$ \hspace{1.65in} $(b)$}
  \caption{{\small A compact Levi graph representation of \emph{policing protest} as a
      frame-action hypergraph  for $(a)$ the  transition function, and
      $(b)$ the reward function at site 0.  Variables $x$ and $x'$ represent the start and end intensities of the protest at site 0 and the action shows the location of the two police troops. As two police troops are sufficient to de-escalate any protest, the contexts in which both troops are at site 0 are independent of the actions of other agents. All other contexts depend on the agents choosing to protest at site 0 only.}}
  \label{fig:Policing}
\end{figure}

We implemented a simple and systematic I-POMDP solving technique that computes reachable beliefs over the finite horizon and then calculates the optimal value at the root node using the Bellman equation for the Many-Agents \ipomdp{} framework. We evaluate its performance in the aforementioned non-cooperative policing protest scenario ($|S|=27, |A_0| = 9, |A_j|=4, |O_j|=8, |O_i|=8$). We model the other agents as \pomdp{}s and solve them using bounded policy iteration~\cite{Poupart03:Bounded}, representing the models as finite state controllers. This representation enables us to have a compact model space. We set the maximum planning horizon to 4 throughout the experiments. The frame-action hypergraphs are encoded into the transition, observation and reward functions of the Many-Agent \ipomdp{} (Fig.~\ref{fig:Policing}). All computations are carried out on a RHEL platform with 2.80 GHz processor and 4 GB memory. 

To evaluate the computational gain obtained by exploiting problem structures, we implemented a solution algorithm similar to the one described earlier that does not exploit any problem structure. A comparison of the Many-Agent \ipomdp{} with the original \ipomdp{} yields two important results: ($i$) When there are few other agents, the Many-Agent \ipomdp{} provides exactly the same solution as the original \ipomdp{} but with reduced running times by exploiting the problem structure. ($ii$) Many-Agent \ipomdp{} scales to larger agent populations, from 100 to 1,000+, and the new framework delivers promising results within reasonable time. 

\begin{table}[!ht]
\small{
\begin{tabular}{|c|c|c|c|c|}
\hline
{\bf Protestors} & {\bf H} & {\bf I-POMDP} & {\bf Many-Agent} & {\bf Exp. Value} \\
\hline
\multirow{2}{*}{2} & 2 & 1 s & 0.55 s & 77.42\\
& 3 & 19 s & 17 s & 222.42\\
\hline
\multirow{2}{*}{3} & 2 & 3 s & 0.56 s & 77.34\\
& 3 & 38 s & 17 s & 222.32\\
\hline
\multirow{2}{*}{4} & 2 & 39 s & 0.57 s & 76.96\\
& 3 & 223 s & 17 s & 221.87\\
\hline
\multirow{2}{*}{5} & 2 & 603 s & 0.60 s & 76.88\\
& 3 & 2,480 s & 18 s & 221.77\\
\hline
\end{tabular}}
\caption{{\small Comparison between traditional I-POMDP and Many-Agent I-POMDP both following same solution approach of computing a reachability tree and performing backup.}}
\label{tbl:smallProblem}
\end{table}

In the first setting, we consider up to 5 protestors with different frames. As shown in Table~\ref{tbl:smallProblem}, both the traditional and the Many-Agent \ipomdp{} produce policies with the same expected value. However, as the Many-Agent \ipomdp{} losslessly projects joint actions to configurations, it requires much less running time. 

\begin{figure}[ht]
  \centerline{\includegraphics[width=2.45in]{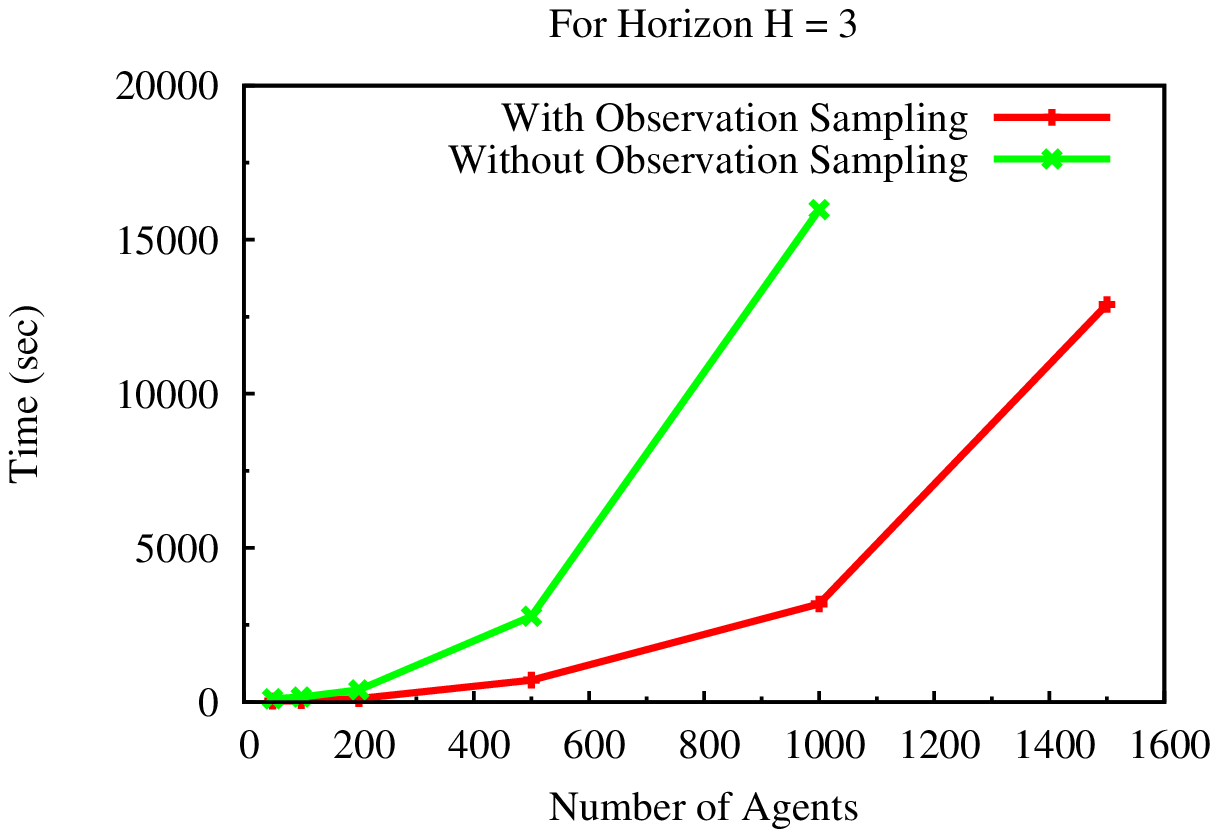}}
  \centerline{{\small $(a)$}}
  \centerline{\includegraphics[width=2.45in]{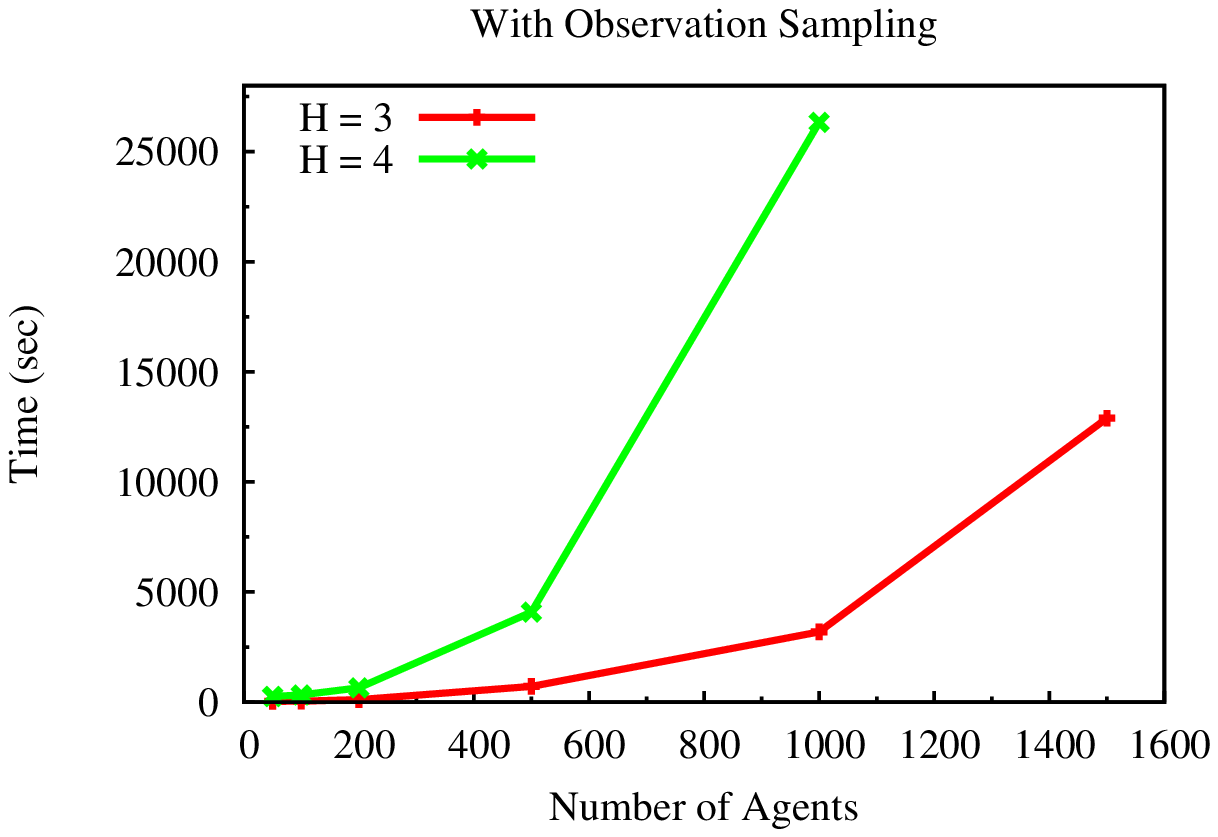}}
  \centerline{{\small $(b)$}}
  \caption{{\small $(a)$ The computational gain obtained by observation sampling. $(b)$ Performance of Many-Agent \ipomdp{} with observation sampling for horizon 3 and 4. The time required to solve a problem is polynomial in the number of agents.}}
  \label{fig:Graphs}
\end{figure}

Our  second  setting considers a large  number of protestors, for which the  traditional \ipomdp{} does not scale.  Instead, we first
scale up the exact solution  method using Many-Agent \ipomdp{} to deal
with a few hundreds of  other agents. Although the exploitation of the
problem structures  reduces the  curse of dimensionality  that plagues
\ipomdp{}s, the  curse of history  is unaffected by such  approaches. To
mitigate  the curse  of  history  we use  the  well-known  observation
sampling method~\cite{Doshi09:Monte}, which  allows us to scale to over
1,000 agents in a reasonable time of 4.5 hours as we show in Fig.~\ref{fig:Graphs}$(a)$. This increases to about 7 hours if we extend the horizon to 4 as shown in Fig.~\ref{fig:Graphs}$(b)$.

\section{Conclusion}

The  key  contribution of  the  Many-Agent \ipomdp{}  is  its
scalability  beyond   1,000  agents  by   exploiting   problem
structures.  We formalize  widely  existing  problem structures -- frame-action  anonymity and context-specific independence -- and
encode it as frame-action hypergraphs.  Other real-world examples
exhibiting such problem structure are found in economics where the value of
an asset depends on the  number of agents vying to acquire it and
their financial standing (frame),  in real estate where the value of
a property depends on its  demand, the valuations of neighboring
properties as well as the economic status of the neighbors  because   an  upscale  neighborhood is  desirable.
Compared  to the previous best approach~\cite{Sonu14:Scalable},  which 
scales to an extension of the simple tiger problem involving 5 agents only,
the presented framework is far more scalable in terms of number of agents.
Our future  work includes exploring other types  of problem structures
and  developing  approximation  algorithms  for this 
\ipomdp{}. An integration with existing multiagent simulation platforms to illustrate
the behavior of agent populations may be interesting.

\section*{Acknowledgements}

This research is supported in part by a NSF CAREER grant, IIS-0845036, and a grant 
from ONR, N000141310870. We thank Brenda Ng for valuable feedback that led to improvements 
in the paper.


\clearpage
\newpage

\section{Appendix}
\underline{\textbf{Factored Belief Update:}}
\begin{align*}
& b_{0,l}(s^{t+1}, m_1^{t+1}, \ldots, m_n^{t+1}) \\
&= Pr(s^{t+1}, m_1^{t+1}, \ldots, m_n^{t+1}|b_0^t, a_0^t, \omega_0^{t+1})\\
&= Pr(s^{t+1}|b_0^t, a_0^t, \omega_0^{t+1}) \times Pr(m_1^{t+1}|s^{t+1}, b_0^t, a_0^t, \omega_0^{t+1})\\
&~~~ \times \cdots \times Pr(m_n^{t+1}|s^{t+1}, m_1^{t+1}, \ldots, m_{n-1}^{t+1} b_0^t, a_0^t, \omega_0^{t+1})\\
\end{align*}

\underline{\textbf{Derivation of equation 5:}}\\
Starting with equation 3, we have:
\begin{align*}
& Pr(s^{t+1}| b_{0,l}^t, a_0^t, \omega_0^{t+1}) \propto Pr(s^{t+1}, \omega_0^{t+1}| b_{0,l}^t, a_0^t) \nonumber \\
& = Pr(s^{t+1}| b_{0,l}^t, a_0^t) Pr(\omega_0^{t+1}|s^{t+1}, b_{0,l}^t, a_0^t)
\end{align*}

The update term for physical states may be represented as a product of its factors such that for any factor $X_k$:
\begin{align*}
& Pr(x_k^{t+1}| b_{0,l}^t, a_0^t) = \sum \limits_{s^t} b_{0,l}^t(s^t) \sum \limits_{\m_{-0}^t} b_{0,l}^t(\m_{-0}^t|s^t) \times  \nonumber\\
&\sum \limits_{\aaa_{-0}^t} Pr(\aaa_{-0}^t|\m_{-0}^t) ~T_0(x_k^t, \langle a_0^t, \aaa_{-0}^t \rangle, x_k^{t+1}) \nonumber\\
\end{align*}
where $b_{0,l}^t(\m_{-0}^t|s^t) = b_{0,l}^t(m_{1}^t|s^t)\times \ldots \times b_{0,l}^t(m_{N}^t|s^t)$, and $Pr(\aaa_{-0}^t|\m_{-0}^t) = Pr(a_{1}^t|m_{1}^t) \times \ldots \times Pr(a_{N}^t|m_{N}^t)$.

We introduce a projection function $\delta^{\nu(\psi)}$ that maps joint actions to the corresponding frame-action configurations as defined in definition 4. 
Formally $\delta^{\nu(\psi)}:\aaa \rightarrow \mathbf{C}^{\nu(\psi)}$, where $\mathbf{C}^{\nu(\psi)}$ is the set of all possible configurations such that for all agents $j$ with frame $\hat{\theta}$, $\C(a,\hat{\theta}) = |\{j: a_j=a, \hat{\theta}_j = \hat{\theta}, (a_j, \hat{\theta}) \in \nu(\psi)\}|$.

Next we partition the set of joint action of all other agents $\bf{A}_{-0}$ into smaller subsets $\bf{A}_{-0}^1,~\ldots,~\bf{A}_{-0}^{|\mathbf{C}^{\nu(\psi)}|}$ such that the projection function $\delta^{\nu(\psi)}$ maps all joint actions belonging to any given partition $\bf{A}_{-0}^c$ to the same value configuration. Hence, we may rewrite the above equation as:

\begin{align*}
& Pr(x_k^{t+1}| b_{0,l}^t, a_0^t) = \sum \limits_{s^t} b_{0,l}^t(s^t) \sum \limits_{c=1}^{|\mathbf{C}^{\nu(x_k^t,a_0^t,x_k^{t+1})}|}  \sum \limits_{\m_{-0}^t} \nonumber\\
& b_{0,l}^t(\m_{-0}^t|s^t) \sum \limits_{\aaa_{-0}^t \in \bf{A}_{-0}^c} Pr(\aaa_{-0}^t|\m_{-0}^t) \nonumber\\
&~T_0(x_k^t, \langle a_0^t, \aaa_{-0}^t \rangle, x_k^{t+1}) \nonumber\\
\end{align*}

Under frame-action anonymity and frame-action independence, for all joint actions $\aaa_{-0}^t \in \bf{A}_{-0}^c$ $T_0(x_k^t, \langle a_0^t, \aaa_{-0}^t \rangle, x_k^{t+1})$ 
$=$ $T_0(x_k^t, a_0^t, \C^{\nu(x^t,a_0^t,x^{t+1})}, x_k^{t+1})$, where $\C^{\nu(x_k^t,a_0^t,x_k^{t+1})}$ 
$=$ $\delta^{\nu(x_k^t,a_0^t,x_k^{t+1})}(\aaa_{-0}^t)$.

\begin{align*}
& Pr(x_k^{t+1}| b_{0,l}^t, a_0^t) 
= \sum \limits_{s^t} b_{0,l}^t(s^t) \sum \limits_{c=1}^{|\mathbf{C}^{\nu(x_k^t,a_0^t,x_k^{t+1})}|}  \sum \limits_{\m_{-0}^t}  \nonumber\\
& b_{0,l}^t(\m_{-0}^t|s^t) \sum \limits_{\aaa_{-0}^t \in \bf{A}_{-0}^c} Pr(\aaa_{-0}^t|\m_{-0}^t) \nonumber\\
& T_0(x_k^t,  a_0^t, \C^{\nu(x_k^t,a_0^t,x_k^{t+1})}, x_k^{t+1}) \nonumber\\
\end{align*}

The cumulative probability of joint actions  mapping to the same configuration, $\sum \limits_{\m_{-0}^t} b_{0,l}^t(\m_{-0}^t|s^t)$ $\sum \limits_{\aaa_{-0}^t \in \bf{A}_{-0}^c} Pr(\aaa_{-0}^t|\m_{-0}^t)$,  is computed tractably using algorithm 1 as $Pr(\C^{\nu(x_k^t,a_0^t,x_k^{t+1})}|b_{0,l}(M_1^t|s^t), \ldots b_{0,l}(M_n^t|s^t)) $. Hence, the equation becomes:
\begin{align*}
& Pr(x_k^{t+1}| b_{0,l}^t, a_0^t) = \sum \limits_{s^t} b_{0,l}^t(s^t) \sum \limits_{\C^{\nu(x_k^t,a_0^t,x_k^{t+1})}} Pr(\C^{\nu(x_k^t,a_0^t,x_k^{t+1})}| \nonumber\\
&b_{0,l}(M_1^t|s^t), \ldots b_{0,l}(M_n^t|s^t))~\times ~T_0(x^t, a_0^t, \C^{\nu(x_k^t,a_0^t,x_k^{t+1})}, x^{t+1}) \nonumber\\
\end{align*}

Similarly, the observation probability may also be obtained in a factored form and $\C^{\nu(s^{t+1},a_0^t,\omega_0^{t+1})} = \delta^{\nu(s^{t+1},a_0^{t},\omega_0^{t+1})}(\aaa_{-0}^t)$ may be substitued instead of the joint action.

\begin{align*}
& Pr(\omega_0^{t+1}|s^{t+1}, b_{0,l}^t, a_0^t) = \sum \limits_{\s^t} b_{0,l}^t(\s^t) \prod \limits_{k=1}^{K}  \sum_{\C^{\nu(x_k^{t+1},a_0^t,\omega_0^{t+1})}}  \nonumber \\
& Pr(\C^{\nu(x_k^{t+1},a_0^t,\omega_0^{t+1})}| b_{0,l}^t(M_{1,l-1}|\s^t), \ldots, b_{0,l}^t(M_{N,l-1}|\s^t))  \nonumber\\
& \OO_0(x_k^{t+1},a_0^t,\C^{\nu(x_k^{t+1}, a_0^t,\omega_0^{t+1})}, \omega_0^{t+1})
\end{align*}

Therefore, we may rewrite equation 3 as follows:
\begin{align}
& Pr(s^{t+1}| b_{0,l}^t, a_0^t, \omega_0^{t+1}) \propto \bigg \{ \sum \limits_{\s^t} b_{0,l}^t(\s^t) \prod \limits_{k=1}^{K}  \sum_{\C^{\nu(x_k^{t+1},a_0^t,\omega_0^{t+1})}} \nonumber\\
& Pr(\C^{\nu(x_k^{t+1},a_0^t,\omega_0^{t+1})}| b_{0,l}^t(M_{1,l-1}|\s^t), \ldots, b_{0,l}^t(M_{N,l-1}|\s^t)) \nonumber\\
& \OO_0(x_k^{t+1},a_0^t,\C^{\nu(x_k^{t+1}, a_0^t,\omega_0^{t+1})}, \omega_0^{t+1}) \bigg \}  \times \bigg \{ \sum \limits_{\s^t} b_{0,l}^t(\s^t) \prod \limits_{k=1}^{K} \nonumber \\
& \sum_{\C^{\nu(x_k^t,a_0^t,x_k^{t+1})}}
Pr(\C^{\nu(x_k^t,a_0^t,x_k^{t+1})}|b_{0,l}^t(M_{1,l-1}|\s^t), \ldots, \nonumber\\
&  b_{0,l}^t(M_{N,l-1}|\s^t))\T_0(x_k^t, \C^{\nu(x_k^t,a_0^t, x_k^{t+1})}, x_k^{t+1}) \bigg \}
\label{eqn:update_state_config_derivation}
\end{align}

\underline{\textbf{Derivation of equation 6:}}\\
The belief update over the models of agent $j$ shown in equation 4 can be rewritten as follows:
\begin{align*}
& Pr(m_{j,l-1}^{t+1}|s^{t+1}, m_{j+1,l-1}^{t+1}, \ldots, m_{N,l-1}^{t+1}, b_{0,l}^t, a_0^t, \omega_{0}^{t+1}) \nonumber\\
& = \sum_{s^t} b_0^t(s^t) \sum_{\m_{1,l-1}^t} b_{0,l}^t(m_{1,l-1}^t|s^t) \sum_{a_{1}^t} Pr(a_1^t|m_{1,l-1}^t) \nonumber\\
& \ldots \sum_{\m_{j-1,l-1}^t} b_{0,l}^t(m_{j-1,l-1}^t|s^t) \sum_{a_{j-1}^t} Pr(a_{j-1}^t|m_{j-1,l-1}^t) \nonumber\\ 
& \sum_{\m_{j,l-1}^t} b_{0,l}^t(m_{j,l-1}^t|s^t) \sum_{a_{j}^t} Pr(a_j^t|m_{j,l-1}^t) \sum_{\m_{j+1,l-1}^t} \nonumber\\
& b_{0,l}^t(m_{j+1,l-1}^t|s^t) \sum_{a_{j+1}^t} Pr(a_{j+1}^t|m_{j,l-1}^t)~ \ldots ~ \sum_{\m_{N,l-1}^t} \nonumber\\
& b_{0,l}^t(m_{N,l-1}^t|s^t)
  \sum_{a_{N}^t}
 Pr(a_N^t|m_{N,l-1}^t) \sum_{\omega_j^{t+1}}  \nonumber\\ 
& O_j(s^{t+1},\langle a_j, \aaa_{-j}^t \rangle, \omega_j^{t+1})~Pr(m_j^{t+1}|m_j^t, a_j^t, \omega_j^{t+1}) 
\end{align*}

Substituting frame-action configuration as in equation 9, we get:
\begin{align}
& Pr(m_{j,l-1}^{t+1}|s^{t+1}, m_{j+1,l-1}^{t+1}, \ldots, m_{N,l-1}^{t+1}, b_{0,l}^t, a_0^t, \omega_{0}^{t+1}) \nonumber\\
& = \sum_{\s^t} b_0^t(\s^t) \sum_{m_j^t} b_0^t(m_j^t|s^t)  \sum_{a_j^t} Pr(a_j^t|m_j^t) \sum_{\C^{\nu(x^{t+1},a_j^t,\omega_{j})}} \nonumber\\
& Pr(\C^{\nu(x^{t+1},a_j^t,\omega_j)}|a_0^t, b_{0,l}^t(M_{1}|\s^t), \ldots, b_{0,l}^t(M_{j-1}|\s^t),\nonumber\\
&   b_{0,l}^t(M_{j+1}|\s^t), \ldots
b_{0,l}^t(m_{N}|\s^t)) \times \sum_{o_j^{t+1}} \OO_j(x^{t+1},a_j^t, \nonumber\\
& \C^{\nu(x^{t+1},\omega_j)},\omega_j^{t+1}) ~Pr(m_j^{t+1}|m_j^t, a_j^t, \omega_j^{t+1}) 
\label{eqn:update_model_config_derivation}
\end{align}

Where the probability over the configurations is computed as in algorithm 1 using belief over models of all other agents except $j$. In the end we add 1 to the count of action $a_0$ in every configuration.

\underline{\textbf{Complexity of belief update:}}\\
For the traditional \ipomdp{} belief update, the complexity of computing equation 3 is $\mathcal{O}(|S|(|M_j^*|)^{N}(|A_j^*|)^{N})$ and that for computing equation 4 is $\mathcal{O}(|S||M_j^*|^N|A_j^*|^N|\Omega_j^*|)$ where $^*$ denotes the maximum cardianlity of a set for any agent. For factored representation, belief update operator invokes equation 3 for each value of all state factors and it invokes equation 4 for each model of each agent $j$ and for all values of updated states. Hence the total complexity of belief update is $\mathcal{O}(K|X^*||S||M_j^*|^N|A_j^*|^N$ $+N|M_j^*||S|^2(|M_j^*|)^{N}|A_j^*|^N|\Omega_j^*|)$.

In equation 5, algorithm is called once for all values of $\s^t$. The two inner summations iterate over all possible configurations over transition and observation contexts. The number of configurations is upper bounded by $\binom{N+|\nu^*|}{|\nu^*|}$ where $|\nu^*|$ is the maximum cardinality of the frame-action neighborhood for any context. Hence the complexity of computing updated belief of state factor $x^{t+1}$ using equation 5 is $\mathcal{O}(|S|K\times\{N|M_j^*||A_j^*||\nu^*|\binom{N+|\nu^*|}{|\nu^*|}$ 
$+\binom{N+|\nu^*|}{|\nu^*|}\})$ (recall the complexity of algorithm 1). Similarly, the complexity of computing updated model probability using equation 6 is  $\mathcal{O}(|S|\times\{N|M_j^*||A_j^*||\nu^*|\binom{N+|\nu^*|}{|\nu^*|}+|\Omega_j^*|$ $\binom{N+|\nu^*|}{|\nu^*|}\})$. These complexity terms are polynomial in $N$ for small values of $|\nu^*|$ as opposed to exponential in $N$ as in equations 3 and 4. The overall complexity of belief update is also polynomial in $N$.

\underline{\textbf{Proof of Proposition 1:}}\\
The expected reward of agent $0$ is obtained as the sum of reward factors.
\begin{align*}
& ER_0(b_{0,l}^t, a_0^t) =  \sum_{\s^t} b_{0,l}^t(s^t) \bigg ( 
 \sum_{k=1}^K \sum_{\m_{-0}^t}
b_{0,l}^t(m_1^t|s^t) \times \ldots \times  \nonumber \\
&  b_{0,l}^t(m_N^t|s^t) \times \sum \limits_{\aaa_{-0}^t} Pr(a_1^t|m_{1,l-1}^t) \times \ldots  \times Pr(a_N^t|m_{N,l-1}^t) \nonumber\\
& R_0(x_k^t,\langle a_0^t, \aaa_{-0}^t \rangle)  \bigg )
\label{eqn:exp_reward_derivation}
\end{align*}

Complexity of computing expected reward using the above equation is $\mathcal{O}(|S|K(|M_j^*|)^N(|A_j^*|)^N)$.
Equation 8 is derived similarly to the belief update by substituting distribution over frame-action configurations for distributions over joint models and joint actions. This combined with the proofs for Eqs. 5 and 6 allow us to obtain Eq.7 from the Bellman equation of the original \ipomdp{}. 

The complexity of computing expected reward using equation 8 is\\
$\mathcal{O}(|S|K\{N|M_j^*||A_j^*||\nu^*|+1\}\binom{N+|\nu^*|}{|\nu^*|})$ which is again polynomial in $N$ for low values of $|\nu^*|$.

\end{document}